\documentclass[12pt]{achemso}
\usepackage[version=3]{mhchem} 
\usepackage{gensymb}
\usepackage{ragged2e}
\usepackage{upgreek}

\usepackage{geometry}
\usepackage{caption}
\usepackage{abstract}

\newcommand\bapbbr {(BA)$_{2}$PbBr$_{4}$}
\newcommand\dd {\mathrm{d}}
\newcommand\purcell {P_{\rm f}}

\usepackage{etoolbox}

\makeatletter
\renewcommand*{\acs@author@fnsymbol@symbol}[1]{
    \ifcase #1 *\or
    1\or
    2\or
    3\or
    4\or
    5\or
    6\or
    7\or
    8\or
    9\or
    10
    \fi
}
        
\renewcommand*\acs@contact@details{
    {\sffamily \fontfamily{ptm}\fontsize{12}{12}\selectfont *\,E-mail: \acs@email@list }%
    \acs@number@list
}

\renewcommand*\acs@contact@details{
    {\sffamily \fontfamily{ptm}\fontsize{12}{12}\selectfont *\,E-mail: muhammad.birowosuto@port.lukasiewicz.gov.pl; liangjie.wong@ntu.edu.sg }%
    \acs@number@list
}

\patchcmd{\acs@address@list@auxii}
{\acs@author@fnsymbol{\acs@affil@marker@cnt}}
{\textsuperscript{\acs@author@fnsymbol{\acs@affil@marker@cnt}}}
{}{}

\patchcmd{\acs@address@list@auxii}
{{\acs@author@fnsymbol{\acs@affil@marker@cnt}\@nameuse{@altaffil@\@roman\@tempcnta}\par}}
{{\textsuperscript{\acs@author@fnsymbol{\acs@affil@marker@cnt}}\@nameuse{@altaffil@\@roman\@tempcnta}\par}}
{}{}        

\makeatother  



\usepackage{mathptmx}

\author{\fontfamily{ptm}\fontsize{12}{12}\selectfont Wenzheng Ye}
\affiliation[EEE]{\fontsize{12}{12}\selectfont \rm School of Electrical and Electronic Engineering, Nanyang Technological University, Singapore 639798, Singapore}\alsoaffiliation[CINTRA]
{CINTRA UMI CNRS/NTU/THALES 3288, Research Techno Plaza, 50 Nanyang Drive, Border X Block, Level 6, Singapore 637553, Singapore}
\author{Zhihua Yong}
\affiliation[EEE]{\fontsize{12}{12}\selectfont \rm School of Electrical and Electronic Engineering, Nanyang Technological University, Singapore 639798, Singapore}\alsoaffiliation[CINTRA]
{CINTRA UMI CNRS/NTU/THALES 3288, Research Techno Plaza, 50 Nanyang Drive, Border X Block, Level 6, Singapore 637553, Singapore}
\author{Michael Go}
\affiliation[EEE]{\fontsize{12}{12}\selectfont \rm School of Electrical and Electronic Engineering, Nanyang Technological University, Singapore 639798, Singapore}\alsoaffiliation[CINTRA]
{CINTRA UMI CNRS/NTU/THALES 3288, Research Techno Plaza, 50 Nanyang Drive, Border X Block, Level 6, Singapore 637553, Singapore}
\author{Dominik Kowal}
\affiliation[Lukasiewicz]{\fontsize{12}{12}\selectfont \rm Łukasiewicz Research Network-PORT Polish Center for Technology Development, Stabłowicka 147, 54-066 Wrocław, Poland}
\author{Francesco Maddalena}
\affiliation[EEE]{\fontsize{12}{12}\selectfont \rm School of Electrical and Electronic Engineering, Nanyang Technological University, Singapore 639798, Singapore}\alsoaffiliation[CINTRA]
{CINTRA UMI CNRS/NTU/THALES 3288, Research Techno Plaza, 50 Nanyang Drive, Border X Block, Level 6, Singapore 637553, Singapore}
\author{Liliana Tjahjana}
\affiliation[EEE]{\fontsize{12}{12}\selectfont \rm School of Electrical and Electronic Engineering, Nanyang Technological University, Singapore 639798, Singapore}\alsoaffiliation[CINTRA]
{CINTRA UMI CNRS/NTU/THALES 3288, Research Techno Plaza, 50 Nanyang Drive, Border X Block, Level 6, Singapore 637553, Singapore}
\author{Wang Hong}
\affiliation[EEE]{\fontsize{12}{12}\selectfont \rm School of Electrical and Electronic Engineering, Nanyang Technological University, Singapore 639798, Singapore}\alsoaffiliation[CINTRA]
{CINTRA UMI CNRS/NTU/THALES 3288, Research Techno Plaza, 50 Nanyang Drive, Border X Block, Level 6, Singapore 637553, Singapore}
\author{Arramel Arramel}
\affiliation[Nano]{\fontsize{12}{12}\selectfont \rm Nano Center Indonesia, Jalan Raya PUSPIPTEK, South Tangerang, Banten, 15314, Indonesia}
\author{Christophe Dujardin}
\affiliation[CNRS]{\fontsize{12}{12}\selectfont \rm Universite Claunde Bernard Lyon 1, Institut Lumière Matière, UMR 5306 CNRS, Villeurbanne F-69622, France}\alsoaffiliation[IUF] {Institut Universitaire de France, 1 Rue Descartes, Paris, Île-de-France, 75005, France}
\author{Muhammad Danang Birowosuto}
\email{muhammad.birowosuto@port.lukasiewicz.gov.pl}
\affiliation[Lukasiewicz]{\fontsize{12}{12}\selectfont \rm Łukasiewicz Research Network-PORT Polish Center for Technology Development, Stabłowicka 147, 54-066 Wrocław, Poland}
\author{Liang Jie Wong}
\email{liangjie.wong@ntu.edu.sg}
\affiliation[EEE]{\fontsize{12}{12}\selectfont \rm School of Electrical and Electronic Engineering, Nanyang Technological University, Singapore 639798, Singapore}
\alsoaffiliation[CINTRA]
{CINTRA UMI CNRS/NTU/THALES 3288, Research Techno Plaza, 50 Nanyang Drive, Border X Block, Level 6, Singapore 637553, Singapore}

\keywords{scintillators, nanophotonics, Purcell effect, spontaneous emission, plasmonics, X-ray imaging}

\title[The Nanoplasmonic Purcell Effect in Ultrafast and High-Light-Yield Perovskite Scintillators]
  {\fontfamily{ptm}\fontsize{12}{12}\selectfont The Nanoplasmonic Purcell Effect in Ultrafast and High-Light-Yield Perovskite Scintillators}

\begin{document}


\noindent
\textbf{The development of X-ray scintillators with ultrahigh light yields and ultrafast response times is a long sought-after goal. In this work, we theoretically predict and experimentally demonstrate a fundamental mechanism that pushes the frontiers of ultrafast X-ray scintillator performance: the use of nanoscale-confined surface plasmon polariton modes to tailor the scintillator response time via the Purcell effect. By incorporating nanoplasmonic materials in scintillator devices, this work predicts over 10-fold enhancement in decay rate and 38\% reduction in time resolution even with only a simple planar design. We experimentally demonstrate the nanoplasmonic Purcell effect using perovskite scintillators, enhancing the light yield by over 120\% to 88 $\pm$ 11 ph/keV, and the decay rate by over 60\% to 2.0 $\pm$ 0.2 ns for the average decay time, and 0.7 $\pm$ 0.1 ns for the ultrafast decay component, in good agreement with the predictions of our theoretical framework. We perform proof-of-concept X-ray imaging experiments using nanoplasmonic scintillators, demonstrating 182\% enhancement in the modulation transfer function at 4 line pairs per millimeter spatial frequency. This work highlights the enormous potential of nanoplasmonics in optimizing ultrafast scintillator devices for applications including time-of-flight X-ray imaging and photon-counting computed tomography.}

\section{\fontsize{12}{20}\selectfont Introduction}

Scintillating materials, which convert ionizing radiation and other high energy particles to visible light, are crucial in a wide range of applications, including medical imaging techniques such as time-of-flight positron emission tomography (TOF PET) \cite{schaart2021physics, schaart2009novel, lecoq2020roadmap, Schaart2020} and photon counting computed tomography (PCCT),\cite{Blaaderen2023,Sar2021} industrial technologies such as security scanning\cite{glodo2017new} and oil drilling exploration,\cite{wibowo2023development, yanagida2013temperature} and scientific measurement processes such as high-energy physics calorimetry and neutrino detection.\cite{cosine2018experiment, araki2005measurement, van2002inorganic} When high energy particles impinge on a scintillator material, a chain of interactions are kicked off, beginning with the excitation of energetic electrons through Compton scattering and the photelectric effect. These energetic electrons subsequently excite electron-hole pairs that propagate to luminescence centers and recombine, emitting visible light via spontaneous emission. 

Scintillator performance is determined by two key properties -- scintillation light yield (as large as possible) and decay time (as small as possible) --, which are crucial in ensuring effective detection capabilities.\cite{Dujardin2018, wang2023, Nikl2006} Approaches to improve scintillator performance include materials development and dopant engineering, \cite{lecoq2016development, birowosuto2016x, yanagida2018inorganic, chen2018all, kakavelakis2020metal, zhang2021reproducible, ma2021highly, heo2018high, zhu2020low} coating scintillators with photonic crystals, \cite{salomoni2018enhancing, knapitsch2014review} implementing fast emitting quantum well structures, \cite{hospodkova2014ingan} and optimizing the associated electronics \cite{birowosuto2006high}. In these works, however, the spontaneous emission rate of the the luminescence centers of scintillators was considered an intrinsic property that remained invariant to changes in the photonic environment of the emitter. Recent breakthroughs,\cite{Kur2020, Ye2022, Cha2022} however, have overturned this thinking by showing that the introduction of nanostructures can provide substantial enhancement in and control over not only the intrinsic decay rate, but even the light yield and emission spectra of scintillator materials. This phenomena arises as a direct manifestation of the Purcell effect. Discovered by Edward Purcell in the 1940s, the Purcell effect refers to the modification in the intrinsic properties of an emitter by its photonic environment, \cite{Pur1946} for instance, through the introduction of nanostructures that alter the radiation modes the emitter couples to. The Purcell effect has been applied to various photonic technologies such as single-photon sources, fluorescence imaging and thermal emitters. \cite{jacob2012broadband,ilic2016tailoring} However, the study of the Purcell effect in nanophotonic scintillators, especially in combination with other advances such as emerging materials and dopant engineering, is still in its infancy, with theoretical studies of Purcell-enhanced nanophotonic scintillators beginning only in 2020, \cite{Kur2020, Ye2022, ye2023, li2023, Shu2023, lahav2022} swiftly followed by experimental demonstrations on enhanced X-ray imaging using two-dimensional photonic crystal scintillators,\cite{Cha2022} and on decay rate enhancement in scintillators with few-hundred-microsecond decay times.\cite{Kurman2023} Notably, all studies of nanophotonic scintillators to date have been limited to photonic crystal structures. Nor has there been an experimental demonstration of the Purcell effect in ultrafast scintillator materials, defined as scintillator materials featuring few-nanosecond (ns) to sub-nanosecond decay times and which are crucial for ultrafast imaging applications. 

Here, we present theoretical studies and experimental demonstrations of the Purcell effect in ultrafast scintillator materials (perovskites), by leveraging the nanoscale confinement of electromagnetic modes in surface plasmon polaritons. We show that the proximity of a plasmonic material, even in a relatively simple planar design, can result in over 10 times enhancements of the scintillator decay rate. The decay rate enhancement in turn leads to better timing performance of scintillators, shortening the time resolution by over 38\%. In effect, we realize these enhancements by engineering the local density of optical states, allowing dipole emitters to outcouple to plasmonic modes, thereby significantly enhancing their decay rate. To accurate simulate our nanophotonic scintillator system, we develop a theoretical framework that takes the existence of surface plasmon polariton modes fully into consideration. In our experiments, we demonstrate enhancements of 120\% and 60\% in light yield per unit thickness and decay rate respectively compared to a bare scintillator system -- this would correspond to a shortening in time resolution of 18\%, which agrees with our theoretical predictions. The layered, planar design that we study in this paper is relevant to applications where large-area, thin film scintillators are commonly employed, such as soft-X-ray imaging and microscopy. More generally, however, our theoretical studies and experimental demonstrations here are proof-of-concepts for the nanoplasmonic Purcell effect in scintillators, paving the way to even more sophisticated designs that leverage surface plasmon polaritons and other types of polaritons.

We perform X-ray imaging experiments using our nanoplasmonic scintillator device to demonstrate significant improvements in the spatial resolution and contrast preservation of the original sample, compared to results obtained with a bare scintillator device. Specifically, we achieve a spatial resolution (line pairs per millimeter) enhancement of 38\% for the modulation transfer function (MTF) at a value of 0.2, and a contrast preservation enhancement of 182\% at a spatial resolution of 4 line pairs per millimeter. Our results pave the way to the use of nanoplasmonic scintillator system in ultrafast imaging systems where high spatial resolution and high contrast are needed, such as X-ray bioimaging and microscopy.

\section{\fontsize{12}{20}\selectfont Results}

\begin{figure}
  \centering
  \includegraphics[width=1\textwidth]{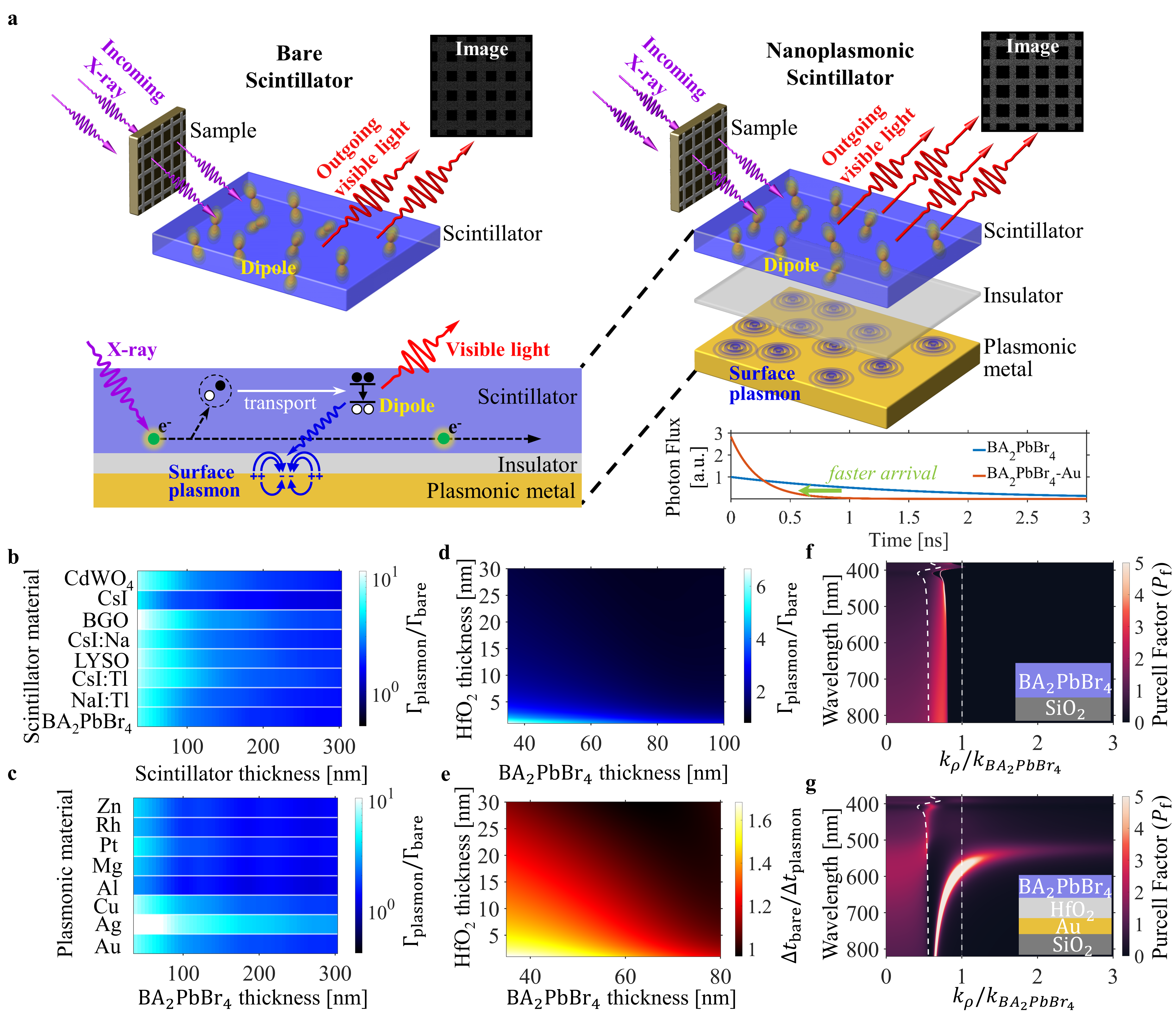}
  \caption{\textbf{Nanoplasmonic Purcell enhancement of perovskite and other scintillators.} a) Schematic illustrations of a bare scintillator system (left) and a nanoplasmonic scintillator system (right), where insulator and plasmonic layers are adjacent to the scintillator layer. When bombarded by X-ray photons (purple pulses), visible photons (red pulses) are generated in nanoplasmonic scintillator system, a process whose decay rate can be significantly enhanced through the existence of surface plasmon polariton modes. The corresponding decrease in time resolution can in turn lead to higher spatial resolution in X-ray imaging. Inset: Illustration of the scintillator process in our nanoplasmonic scintillator device: An incident X-ray photon creates an energetic primary electron (glowing black circle), which subsequently creates secondary electrons (black circle) and holes (white circle). Electrons and holes transport to dopants ions and form electron-hole pairs. Visible light and surface plasmon plariton are emitted through the recombination of electron-hole pairs via spontaneous emission. The spontaneous emission properties can be tailored through the choice of scintillator thickness and plasmonic material, resulting in a Purcell-enhanced photon emission rate (right inset). b) Decay rate enhancement as a function of scintillator material and scintillator thickness, where the insulator is 1-nm-thick HfO$_{2}$ and the plasmonic layer is 70-nm-thick Au. The enhancement is calculated by normalizing the decay rate with that of a bare scinitllator system (with glass substrate) of the same scintillator thickness. c) Decay rate enhancement as a function of plasmonic material and scintillator thickness, where the insulator is 1-nm-thick HfO$_{2}$, the plasmonic layer is 70-nm-thick and the scintillator material is \bapbbr. \bapbbr~ is chosen for this example as it is a cost-effective, solution processable material with high light yield (40 $\pm$ 4 ph/keV) and fast decay time (3.3 $\pm$ 0.3 ns). d, e) Impact of \bapbbr ~thickness and HfO$_{2}$ insulator thickness on the scintillator performance , using 70-nm-thick Au as the plasmonic material. The total decay rate is enhanced and time resolution -- defined as the variance of the first photon arrival time $\Delta t$ -- is shortened when decreasing scintillator thickness and insulator thickness. f, g) Purcell factor $\purcell(\lambda,k_{\rho})$ -- defined as the decay rate enhancement compared to a corresponding dipole in vacuum -- as a function of emission wavelength and the projection of the wavevector onto the plane axis parallel to the layer interfaces, without (f) and with (g) HfO$_{2}$ and Au films. The introduction of the Au film enhances the emission rate of the photons into all types of modes: outgoing radiative modes (left of white dashed lines); guided modes (between two white dashed lines); and surface plasmon polariton modes (right of white dashed lines). Bottom right insets show the configurations considered in the respective panels.}
  \label{fig:boat1}
\end{figure}

\textbf{Figure \ref{fig:boat1}}a contrasts the visualization of a bare scintillator with that of a nanoplasmonic scintillator system, where an externally added plasmonic metal film and an insulator film are adjacent to a scintillator film. For a bare scintillator system, the emission characteristics are solely determined by the intrinsic properties of the scintillator's luminescence centers. However, the situation is dramatically different when we insert a plasmonic thin film next to the scintillator. \cite{CPR78,Ford1984} The emission rate and intensities are then modified by the Purcell effect, which arises through the existence of surface plasmon polariton modes that alter the local density of states from that of the bare scintillator configuration. Enhancements in both light yield and emission rate can benefit many applications, in particular imaging systems, by enhancing the spatial resolution and contrast of ultrafast X-ray imaging. The inset of Figure 1a illustrates the physical processes taking place in the nanoplasmonic scintillator device. High energy particles (X-rays in this work) are converted into hot electrons (glowing circle) through processes such as photoelectric effect and Compton scattering when they pass through scintillator material. These high energy electrons further excite multiple secondary electrons and holes with varying positions and orientations. Electrons and holes are then transported to dopant ions where they form electron-hole pairs (dipole emitters). These electron-hole pairs eventually relax, emitting visible photons. The dipole emission couples primarily into three types of modes: the outgoing radiative modes, guided modes that propagates along the lateral dimensions of the multilayer structure, and surface plasmon polariton modes. The existence of surface plasmon polariton modes significantly enhances the decay rate of dipole emitters in nanoplasmonic scintillator system, particularly for those near the scintillator-metal interface and oriented perpendicular to the surface,\cite{Drx1970,CPR78,Ford1984, akselrod2014probing} as described in detail in Supplementary Information (SI) Section I. To accurately model nanoplasmonic scintillators, we develop a theoretical framework based on our derivations in Ref \cite{Ye2022} with a key difference that is crucial for accuracy in the nanoplasmonic scintillator system that we consider: Our model takes into consideration the fact that dipoles can decay into surface plasmon polariton modes whose in-plane wavenumber is larger than the wavenumber in the scintillator material, i.e.,

\begin{equation}
    \Gamma(\textbf{r}, \lambda) \propto \int_{0}^{\infty} \dd k_{\rho} \frac{k_\rho}{k_{mz} k_m} \purcell(\textbf{r}, \lambda, k_{\rho}).
\end{equation}

\begin{equation}
    \purcell(\textbf{r}, \lambda, k_{\rho}) = \frac{\Gamma(\textbf{r}, \lambda, k_{\rho})}{\Gamma_0(\lambda)}.
\end{equation}
where $\Gamma$, $\purcell$, $\textbf{r}$, $\lambda$, $k_{\rho}$, $k_{mz}$, $k_m$ are the decay rate, the Purcell factor, dipole position, emission wavelength, emission in-plane wavenumber, emission wavenumber in z direction, and emission wavenumber in scintillator, respectively. $\Gamma_0(\lambda)$ is the total decay rate of a dipole radiating at wavelength $\lambda$ in the free space. It is noteworthy that the uppermost integral limit in Equation 1 is $\infty$, indicating that even propagation vectors that fall outside the light cone, and which therefore correspond to polaritons, are fully taken into consideration in our theory. Full details of our theoretical framework are presented in the Methods and SI Section I.

Scintillators can be broadly classified into inorganic and organic scintillators. Inorganic scintillators, such as lanthanide-doped scintillators, typically exhibit high light yield ($>$ 10 ph/keV) and extended decay times ($>$ 10 ns). Organic scintillators typically feature lower light yields ($<$ 10 ph/keV) and faster decay times ($<$ 5 ns).\cite{Maddalena2019rev} Organic scintillators may be disadvantaged by their relatively low mass densities and small effective atomic numbers, which tend to restrict their suitability for specific high-energy applications.\cite{Dujardin2018} In recent years, solution-processable and low-temperature-growth perovskite scintillators have attracted significant interest.\cite{birowosuto2016x,Chen2018nat} These scintillators were initially inspired by the extensive exploration of perovskite photovoltaics over a decade ago.\cite{Dou2015, khalfin2019advances, bekenstein2015highly}  Perovskite scintillators, can be up to 50 times less expensive to manufacture compared to inorganic lanthanide-based scintillators.\cite{Maddalena2019rev} Furthermore, certain perovskite scintillators are known for both high light yield ($>$ 20 ph/keV) and fast decay times ($<$ 5 ns). Compared to organic scintillators, perovskite scintillators typically possess 3 to 6 times larger mass density and effective atomic numbers.\cite{Xie2020,Maddalena2021} These unique properties of perovskite scintillators thus make them highly promising candidates for static X-ray imaging applications,\cite{Chen2018nat} as well as time-resolved diagnostic tools like time-of-flight positron emission tomography and photon-counting computed tomography. \cite{Blaaderen2023,Wibowo2023,Cala2022,Sar2021}.

Figure \ref{fig:boat1}b shows the ability of plasmonic film to tailor the decay rate of different scintillator materials and scintillator thicknesses with a scintillator-insulator(1 nm HfO$_2$)-metal (70 nm gold (Au)) planar design. As examples, the parameters of eight leading scintillator materials are used in the simulation.\cite{kimble2002scintillation, burachas1996large, grodzicka2012mppc, kubota1988new, mengesha1998light, moszynski2002intrinsic, Lili2021,gektin2017inorganic} By varying the scintillator thickness, the decay rate ($\Gamma$) of the scintillators can be enhanced by over 10 times by the introduction of plasmonic film through Purcell effect, especially for scintillator of small thicknesses. For further theoretical exploration and experimental demonstration, we focus on \bapbbr~  as the scintillator material of choice due to two reasons: Firstly, it is cost-effective and relatively easy to grow because it is a solution processable material, making it promising for commercialization \cite{birowosuto2016x}; Secondly, it features both high light yield (40 $\pm$ 4 ph/keV) and fast decay times (3.3 $\pm$ 0.3 ns), which are critical in optimizing the performance of time-of-flight imaging systems and PCCT.\cite{Xie2020,Maddalena2021} Although the nanoplasmonic BGO system has the largest enhancement (11 times, see Figure 1b), it nevertheless suffers from a slower decay time of about hundreds of nanoseconds, compared to the few-nanosecond decays times of perovskite scintillators. \cite{Maddalena2019rev}. Figure \ref{fig:boat1}c shows the ability of different plasmonic materials to tailor the decay rate of scintillators. The decay rate of \bapbbr~ can be enhanced by all plasmonic materials studied, reaching more than 10 times enhancement compared to a bare scintillator system. Figures \ref{fig:boat1}b, \ref{fig:boat1}c highlight the versatility in choice of plasmonic material to design nanoplasmonic scintillator devices.

Figures \ref{fig:boat1}d, \ref{fig:boat1}e investigate the impact of scintillator (\bapbbr) thickness and insulator (HfO$_2$) thickness on the decay rate ($\Gamma$) and time resolution $\Delta t$, defined as the variance of the first photon arrival time. The scintillator decay time and time resolution are shortened with the decrease of scintillator thickness and insulator thickness, which we expect since the nanoscale confinement of the plasmonic modes implies that a larger proportion of luminescence centers will be coupled to more of these plasmonic modes for thinner scintillators. Although the use of nanoplasmonics in our planar configuration limits applications to the soft X-ray regime where the absorption length is also nanoscale, it is also possible to envision the realization of thicker nanoplasmonic scintillator devices through a multilayer configuration and outcoupling radiated light from the sides. The improved timing performance of nanoplasmonic scintillators is relevant to time-resolved applications, such as ultrafast imaging systems, particularly in soft X-ray regime.\cite{du2018x, zhu2020low, van2023bza, pagano2022new, pichette2013time, berube2016prospects, rossignol2020time, lucchini2020new, gundacker2020experimental, vinogradov2018approximations}

Figures \ref{fig:boat1}f, \ref{fig:boat1}g illustrate the enhancement of the dipole decay rate in an exemplary nanoplasmonic scintillator system (45 nm-thick \bapbbr, 15 nm-thick HfO$_2$, 70-nm-thick Au) using the Purcell factor ($\purcell(\lambda,k_{\rho})$), which represents the enhancement in decay rate relative to the natural decay rate of a dipole emitter in vacuum. In figures \ref{fig:boat1}f and \ref{fig:boat1}g, the Purcell factor shown has been integrated over all dipoles -- isotropically oriented -- in the scintillator material, where $\lambda$ and $k_{\rho}$ respectively denote the emission wavelength and the in-plane wavevector of the dipole emission. The outgoing radiative photons occur in the regime where the in-plane wavevector is smaller than the total reflection wavevector in \bapbbr~ (left of both dashed white lines), where the emission rate of out-coupled radiative photons experiences slight enhancement. Emission with an in-plane wavevector larger than the wavevector in \bapbbr~ (right of both white dashed lines) corresponds to a lossy wave within \bapbbr. Near the \bapbbr-Au interface, this lossy wave manifests as a surface plasmon plariton wave, which is not observable far away from or in the absence of the plasmonic material. This phenomenon contributes to a substantial enhancement in the decay rate of dipoles, especially for those luminescence centers located near the \bapbbr-Au interface, as is shown in Supplementary Figure S2. The scintillator-metal waveguided mode, corresponding to the region between the white dashed lines, also experience enhancements in certain wavelength ranges. Therefore, the introduction of the plasmonic film enhances the emission rate of the dipoles into all decay modes.

\begin{figure}
  \centering
  \includegraphics[width=0.9\textwidth]{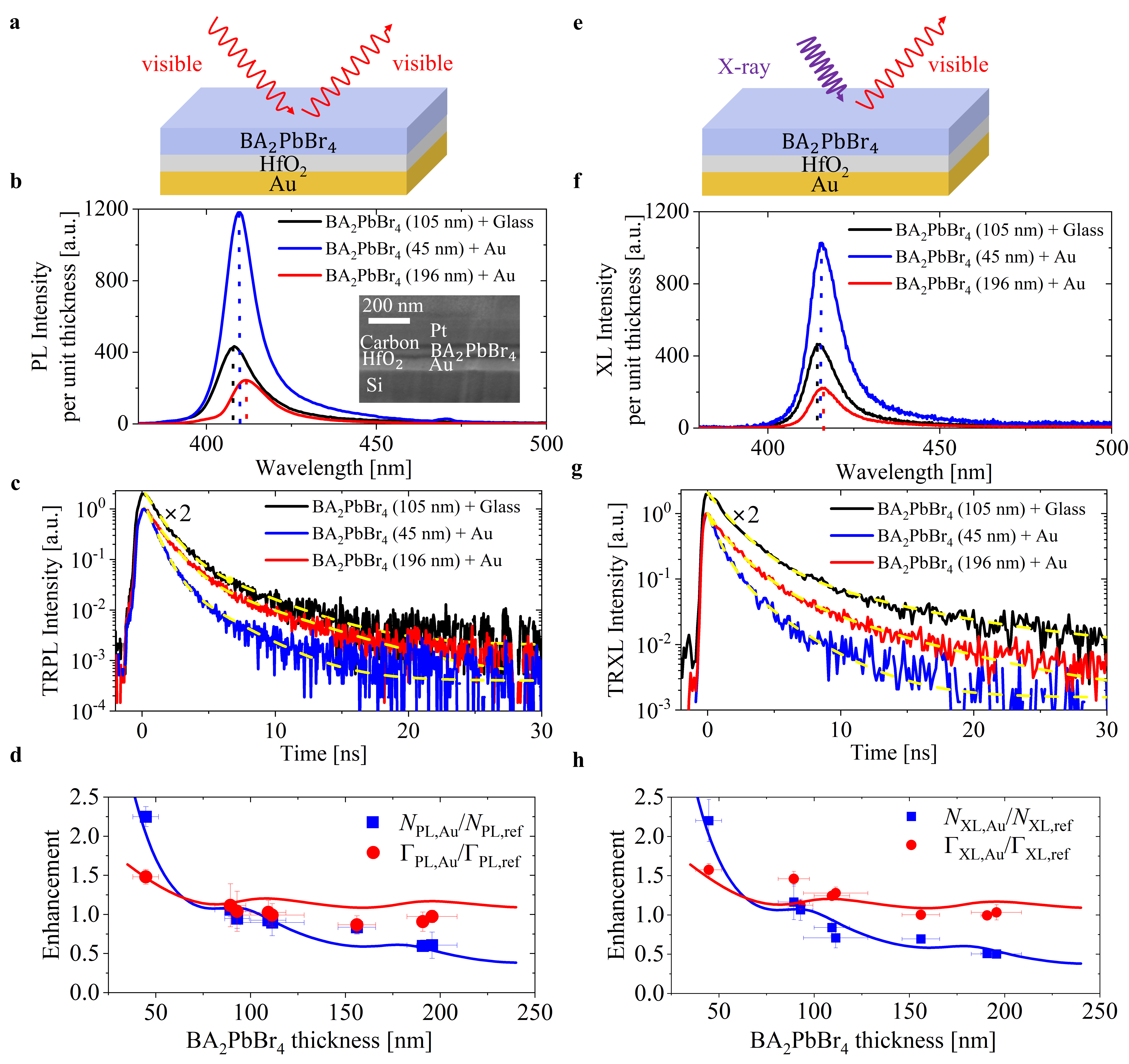}
  \caption{\textbf{Photoluminescence and X-ray excited luminescence experiments demonstrating enhanced decay rate and enhanced intensity per unit thickness in nanoplasmonic scintillators, in good agreement with theoretical predictions.} a) and e) Schematic illustration of the exciation sources and outgoing radiation waves. b) Normalized photoluminescence (PL) spectra, c) time-resolved photoluminescence (TRPL) decay curves, f) normalized X-ray excited luminescence (XL) spectra and g) time-resolved X-ray excited luminescence (TRXL) decay curves for three configurations: 105 nm \bapbbr ~thin film with glass substrate (reference), and two nanoplasmonic \bapbbr~ thin film systems of different scintillator thicknesses (45 nm and 196 nm). The PL spectra and XL spectra show that the nanoplasmonic \bapbbr ~system with the thinner scintillator layer (45 nm \bapbbr) exhibits enhanced intensity per unit thickness, over that of the reference system. Both nanoplasmonic configurations show an enhancement in the decay rate, over the reference scintillator system, as illustrated in TRPL decay curves and TRXL decay curves. Inset: Cross-sectional scanning electron microscope (SEM) image of the nanoplasmonic \bapbbr~ thin-film scintillator. Carbon is added only during SEM characterization to adhere the thin film sample to the platinum (Pt) holder. d) and h) Impact of \bapbbr ~sample thickness on emission intensity per unit thickness ($N$) and decay rate ($\Gamma$). $N_{\rm PL,Au}/N_{\rm PL,ref}$~, $N_{\rm XL,Au}/N_{\rm XL,ref}$, $\Gamma_{\rm PL,Au}/\Gamma_{\rm PL,ref}$ and $\Gamma_{\rm XL,Au}/\Gamma_{\rm XL,ref}$ denote the enhancement in PL intensity per unit thickness, XL intensity per unit thickness, PL decay rate and XL decay rate, respectively. Our experiments thus demonstrate the ability of the nanoplasmonic Purcell effect to enhance the intensity per unit thickness by over a factor of 2, and the decay rate by over a factor of 1.5, for sufficiently thin scintillator films. In the discussion of this work, we emphasize that the need for thin scintillator films does not detract from the usefulness of the nanoplasmonic Purcell effect, as thin scintillator films are useful for energetic particles with short absorption lengths, and can also form thicker scintillators in multilayer configurations.}
  \label{fig:boat2}
\end{figure}

\textbf{Figure \ref{fig:boat2}} experimentally demonstrate the enhancement of the emission intensity and decay rate in the nanoplasmonic scintillator system under ultraviolet (Figures \ref{fig:boat2}a-\ref{fig:boat2}d) and X-ray (Figures \ref{fig:boat2}e-\ref{fig:boat2}h) excitation, using \bapbbr ~as scintillator and Au as plasmonic metal. The choice of these materials were motivated by their availability and our ability to deposit a high-quality Au film with a surface flatness of less than 5 nm at a thickness of 70 nm. \cite{Adamo2020, singh2022cavity, balakrishnan2017cspbbr3, Cortes2022} We investigate three configurations whose results are presented in Figures \ref{fig:boat2}b, \ref{fig:boat2}c, \ref{fig:boat2}f, \ref{fig:boat2}g: (i) a 105 nm \bapbbr ~thin film with a glass substrate (reference), (ii) nanoplasmonic \bapbbr ~thin film systems with two different thicknesses of \bapbbr ~thin film (45 nm and 196 nm). Inset of Figure \ref{fig:boat2}b shows the scanning electron microscope image of a cross section of a sample of metal-perovskite thin film scintillator, specifically a \bapbbr ~thin film (thickness: 45 $\pm$ 4 nm) on a Au thin film (thickness: 70 $\pm$ 7 nm), interfaced by a HfO$_{2}$ ~insulating layer (thickness: 15 $\pm$ 1 nm). The thicknesses of the films are determined using a profilometer. Additional characterizations of the film thicknesses, atomic structures, and compositions of the \bapbbr ~films, obtained through atomic force microscope, X-ray diffraction, and X-ray photoemission spectroscopy measurements, are shown in SI Section II, III, and IV, respectively.

Figures \ref{fig:boat2}b, \ref{fig:boat2}f illustrate the normalized photoluminescence (PL) spectra and normalized X-ray excited luminescence (XL) spectra ($N_{\rm PL}$ and $N_{\rm XL}$: emission intensity per unit thickness) of the three configurations. As a result, the introduction of Au film enables successful tailoring of the PL and XL spectrum, resulting in either an enhancement of 173\% and 120\% (thinner samples), respectively, or suppression of 44 \% and 54\% (thicker samples), respectively, on the peak of normalized intensity. Additionally, the PL spectrum peaks exhibit a shift from 408 nm to 410 nm and 412 nm corresponding to the reference configuration, 45-nm-thick, and 196-nm-thick nanoplasmonic \bapbbr~ samples, respectively. Such shifts can be attributed to the changes of effective refractive indices of the samples.\cite{Birowosuto2010} Similarly to the PL spectra, the emission peaks experience a shift from 413 nm to 416 nm with the introduction of Au film in XL spectra. The additional shift of emission peak from 408 nm to 413 nm between the respective PL and XL spectra of the aforementioned reference sample is attributed to self-absorption, as previously demonstrated in \bapbbr ~thin film.\cite{Maddalena2021} In Figures \ref{fig:boat2}c and \ref{fig:boat2}g, the results of time-resolved photoluminescence (TRPL) and time-resolved X-ray excited luminescence (TRXL) decay curves of the photon emission are presented, respectively, with the decay curve of reference configuration multiplied by 2 for improved clarity. To ensure the successful observation of the Purcell enhancements, the decay time of \bapbbr ~thin film remained stable throughout the 8-day experimental period (SI Section VI). The TRPL decay curves are fitted with two-exponential-decay model and the corresponding fitting parameters ($\tau_{\rm PL,1}$, $\tau_{\rm PL,2}$, and $\bar{\tau}_{\rm PL}$ for the first, second, and averaged decay components, respectively) are shown in Supplementary Table S1. The decay times show a significant decrease in both nanoplasmonic \bapbbr~ systems compared with the reference configuration, particularly obvious for 45-nm-thick \bapbbr~. Meanwhile, the contributions of two decay components remain unchanged: 70\% from $\tau_{\rm PL,1}$ and 30\% from $\tau_{\rm PL,2}$. In both decay components, there is a decrease in $\tau_{\rm PL,1}$ and $\tau_{\rm PL,2}$ from 1.0 $\pm$ 0.1 ns to 0.7 $\pm$ 0.1 ns and from 3.9 $\pm$ 0.4 ns to 2.9 $\pm$ 0.3 ns, respectively, when changing the reference configuration to the 45-nm-thick nanoplasmonic \bapbbr ~system. As a result, $\bar{\tau}_{\rm PL}$ decreases from 1.8 $\pm$ 0.2 ns to 1.3 $\pm$ 0.1 ns for the respective samples. The 0.7 $\pm$ 0.1 ns decay time is another experimental demonstration of the PL decay rate enhancement in sub-ns regime, with previously in photonic crystal \cite{Birowosuto2014} and plasmonic cavity.\cite{Duke2014} The TRXL decay curves are fitted with three decay components as $\tau_{\rm XL,1}$, $\tau_{\rm XL,2}$, $\tau_{\rm XL,3}$, and averaged decay time $\bar{\tau}_{\rm XL}$, see Supplementary Table S2. The presence of slow component $\tau_{\rm XL,3}$ is related to the scintillation mechanism, which is not present during ultraviolet excitation.\cite{Maddalena2021} As a result, the decay times of \bapbbr ~thin film experience a significant reduction due to the presence of the Au film, which can be attributed to the Purcell effect. Specifically, $\tau_{\rm XL,1}$ decreases from 1.5 $\pm$ 0.2 ns to 0.7 $\pm$ 0.1 ns, $\tau_{\rm XL,2}$ decreases from 4.1 $\pm$ 0.4 ns to 1.4 $\pm$ 0.2 ns, $\tau_{\rm XL,3}$ decreases from 7.7 $\pm$ 0.8 ns to 3.4 $\pm$ 0.3 ns, and $\bar{\tau}_{\rm XL}$ decreased from 3.3 $\pm$ 0.3 ns to 2.0 $\pm$ 0.2 ns, respectively, when comparing the decay curves of reference samples with those of the 45-nm-thick nanoplasmonic \bapbbr~ system (see Supplementary Table S2). In the comparison with PL decay time components ($\tau_{\rm PL,1}$ and $\tau_{\rm PL,1}$) and average values $\bar{\tau}_{\rm PL}$, only fastest component is the same while the rests are about 1 ns slower. This is related with the scintillation mechanisms, which $\tau_{\rm XL,1}$ is related with direct electron-hole captures while the slower components, $\tau_{\rm XL,2}$ and $\tau_{\rm XL,3}$ are related with the slower excitons and defects.\cite{Maddalena2019rev} The dipole orientation of this \bapbbr ~thin film is more perpendicular to its surface (SI Section V) supporting the large enhancements of emission intensity per thickness and decay rate in Figures \ref{fig:boat2}b, \ref{fig:boat2}f, and Figures \ref{fig:boat2}c, \ref{fig:boat2}g, respectively. Similar dipole orientation was also observed in other perovskite thin films.\cite{Qin2021} By utilizing this nanoplasmonic perovskite scintillator system, we are able to improve the scintillator performance in the sub-ns regime under both ultraviolet excitation and X-ray excitation.

To further validate the control of spontaneous emission and the shortening of decay times ($\tau_{\rm PL}$ and $\tau_{\rm XL}$) in the nanoplasmonic scintillator system, we investigate the impact of the thickness of \bapbbr ~ on the scintillator performance. We vary the thicknesses of \bapbbr~ thin film from 45 nm $\pm$ 4 nm to 196 nm $\pm$ 20 nm and compare normalized intensity ($N_{\rm PL}$ and $N_{\rm XL}$) and decay rate ($\Gamma_{\rm PL}$ and $\Gamma_{\rm XL}$) of the nanoplasmonic samples (denoted by subscript letter of Au) to those of the reference samples (same with Figures \ref{fig:boat2}b, \ref{fig:boat2}c, \ref{fig:boat2}f, and \ref{fig:boat2}g and denoted with subscript letter of ref). Figure \ref{fig:boat2}d experimentally demonstrate that both $N_{{\rm PL,Au}}/N_{{\rm PL,ref}}$ ~and $\Gamma_{{\rm PL,Au}}/\Gamma_{{\rm PL,ref}}$ ~increase as the \bapbbr~ thickness decreases, particularly for sample thicknesses below 75 nm. The maximum enhancements observed in $N_{\rm PL}$ ~and $\Gamma_{{\rm PL}}$ ~are 130\% $\pm$ 23\% and 40\% $\pm$ 14\%, respectively. Figure \ref{fig:boat2}h experimentally demonstrate that both $N_{\rm XL,Au}/N_{\rm XL,ref}$ ~and $\Gamma_{{\rm XL,Au}}/\Gamma_{{\rm XL,ref}}$ ~increase as the \bapbbr ~thickness decreases, particularly for thicknesses below 100 nm. The maximum enhancements observed in $N_{\rm XL}$ ~and $\Gamma_{\rm XL}$ ~are 120\% $\pm$ 22 $\%$ and 60\% $\pm$ 16 $\%$, respectively. These experimental findings are consistent with the numerical predictions (solid lines). It is hitherto the first experimental demonstration of the XL decay rate enhancement of scintillator with a sub-ns decay time.

\begin{figure} [H]
  \centering
  \includegraphics[width=0.8\textwidth]{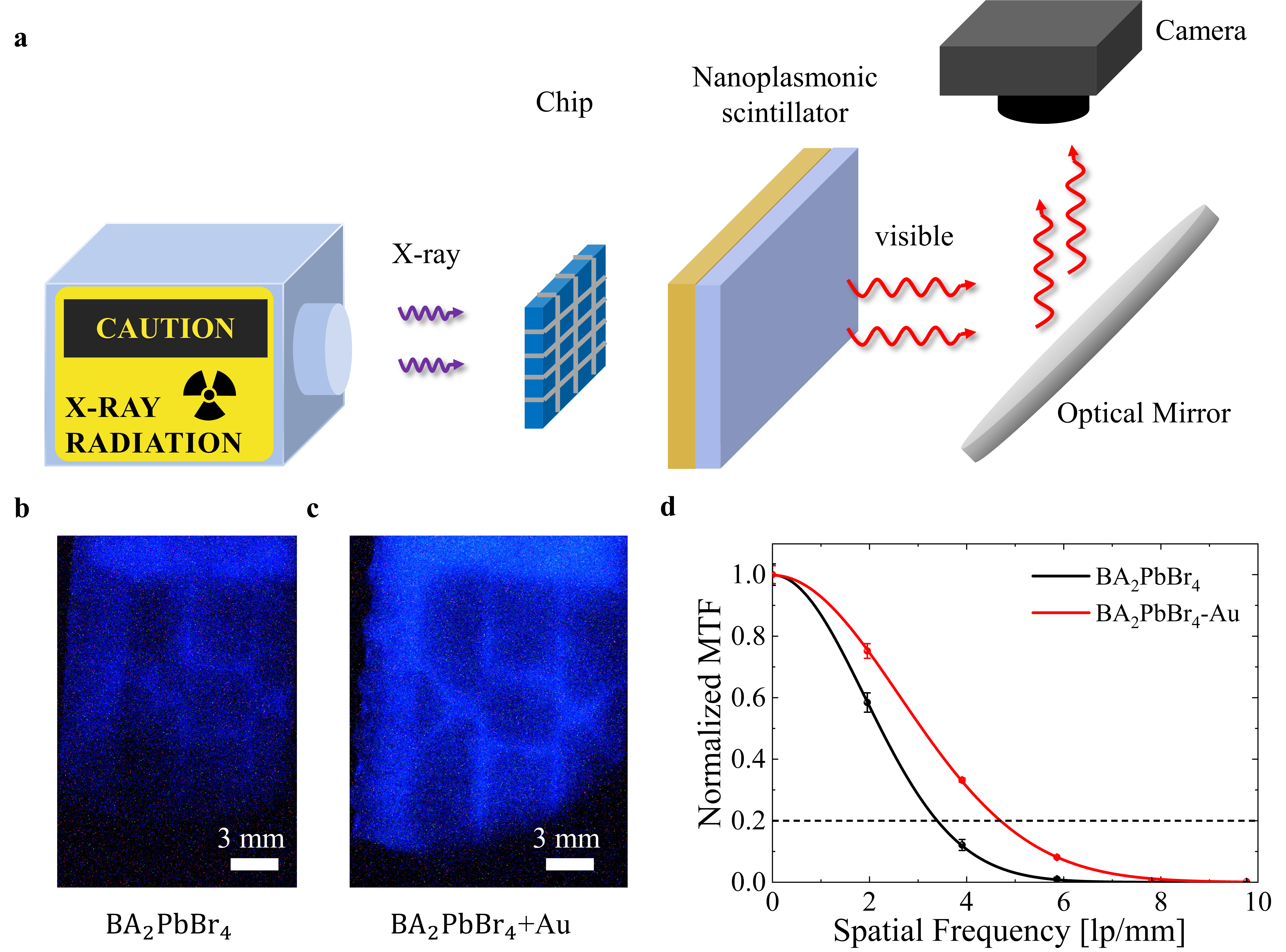}
  \caption{\textbf{Enhancing the spatial resolution of X-ray imaging with a nanoplasmonic scintillator system.} a) Schematic diagram of X-ray imaging system based on our nanoplasmonic scintillator system. b) and c) depict X-ray scanning images of a chip (10 mm) detected by \bapbbr ~ thin film and nanoplasmonic \bapbbr ~system, respectively. The spatial resolution of the image experiences observable enhancements by the introduction of Au film. d) Improving modulation transfer function (MTF) with nanoplasmonic perovskite-Au system. As the spatial frequency of scanning image increases, the nanoplasmonic perovskite-Au system maintains higher MTF performance. The MTF is derived from the line and edge spread functions obtained from the radiographic images of square holes, as shown in SI Section VII.}
  \label{fig:boat3}
\end{figure}

Enhancements in both light yield and decay time have been shown to significantly improve the spatial resolution of X-ray imaging, as demonstrated in Refs. \cite{Kurman2023, Shu2023} Since we have realized a Purcell-enhanced scintillator system facilitated by nanoplasmonic design (depicted in Figure \ref{fig:boat2}). The X-ray imaging results presented in \textbf{Figure \ref{fig:boat3}} further validate the effectiveness of the nanoplasmonic scintillator system in practical use. The schematic set-up is illustrated in Figure \ref{fig:boat3}a. Here, we experimentally demonstrate an enhanced spatial resolution and contrast preservation of X-ray imaging of a chip with nanoplasmonic scintillator system, with observable comparison between Figure \ref{fig:boat3}b and Figure \ref{fig:boat3}c. We use modulation transfer function (MTF) to quantify the imaging contrast preservation of the original object by the detector. The MTFs for both reference and 45-nm-thick nanoplasmonic \bapbbr~ samples are derived from the line and edge spread functions from the images of radiographic modes of square holes in SI Section VII. With the decrease of the contrast preservation of scanning image, the nanoplasmonic scintillator system maintains a higher spatial resolution. For instance, for MTF at a value of 0.2, the spatial frequency for the nanoplasmonic \bapbbr ~sample reaches 4.7 $\pm$ 0.1 line pairs per millimeter (lp/mm), surpassing the 3.4 $\pm$ 0.1 lp/mm achieved by the reference sample. On the other hand, MTF shows an enhancement of 182\%, from 0.11 to 0.31, at 4 lp/mm spatial frequency. The bright observation of X-ray image in thin \bapbbr ~samples is due to their high light yield ($40$ ph/keV),\cite{Xie2020,Maddalena2021, Imagenano1} which confirms the promising practical use of nanoplasmonic scintillator system in imaging devices.

\section{\fontsize{12}{20}\selectfont Discussion}

Plasmonic materials are typically metals or metal-like materials, among which Au and Ag are commonly used. \cite{west2010searching, jablan2009plasmonics} Featuring negative real permittivity, plasmonic materials are widely used in many applications that include biological optical sensing and superlenses. \cite{zhang2012surface} Plasmonic materials are often used to accelerate the dipole decay through the Purcell effect. \cite{tanaka2010multifold, noginov2010controlling, bryant2008mapping} The underlying mechanism of the Purcell enhancement in plasmonic systems is the existence of surface plasmon polariton modes to which dipoles can couple, especially when the dipole is close to plasmonic materials and dipole emission wavelength is close to plasmon resonance peak. \cite{akselrod2014probing, luo2017purcell, kongsuwan2018suppressed} One innovative aspect of our work arises from the fact that Purcell enhancement via nanoplasmonics has not been predicted or demonstrated in scintillator devices. Another innovative aspect lies in the fact that we present an unprecedented demonstration of the Purcell effect in an ultrafast scintillator material, where the intrinsic decay time is already on the order of few-nanoseconds. Ours are thus key experiments in demonstrating the usefulness of the Purcell effect -- by showing that the Purcell effect is highly complementary with other methods of enhancing decay rate, such as material design, and can take a material already at the known limits of timing performance even further. Ultrafast nanoplasmonic scintillators could benefit many ultrafast scintillator applications, such as real-time bioimaging, microscopy and photon-counting computed tomography that needs high scanning speed.\cite{du2018x, zhu2020low, van2023bza} Furthermore, the performance of TOF X-ray imaging systems, where time resolution is the determining factor, can be improved due to the faster decay time.\cite{pagano2022new, pichette2013time, berube2016prospects, rossignol2020time, lucchini2020new, gundacker2020experimental, vinogradov2018approximations, du2018x} More broadly, our findings highlight the relevance of nanophotonics -- the use of nanostructures to mold the flow of light -- to X-ray and free electron science, with nanophotonics having already been shown to enhance the properties of free electron-driven emission at X-rays and other wavelengths.\cite{tan2021space, lu2023smith, nussupbekov2021enhanced, hu2021free, lin2018controlling, lin2021brewster, huang2022enhanced, wong2016towards, huang2023quantum, yang2018maximal, shim2019fundamental}

While the nanoplasmonic Purcell effect is capable of substantially enhancing decay rate and time resolution, it is important to take into consideration the intended application in designing the scintillator device. Due to the nanoscale range of the tightly confined nanoplasmonic modes, a single metal-scintillator interface can only provide Purcell enhancement to the luminescence centers in the nanoscale vicinity of the interface. As such, single-scintillator-layer devices are most suitable for scintillators with a thin active layer. For example, a single-scintillator-layer device could function as a soft X-ray detector, since the absorption length of soft X-rays can be on the order of 10-100 nanometers. \cite{chantler1995theoretical, chantler2000detailed} An example for applications of ultrafast soft X-ray detectors include time-resolved X-ray microscopy in the water window regime for biological imaging of living specimens. \cite{toci2019ingan, nakazato2011evaluation, chen2022nanoscale, yu2019significant, jiang2021recent, fan2015x, li2018soft, naczynski2015x, jiang2020soft} To harness nanoplasmonic technology for scintillator detection of high energy particles such as hard X-rays, one can consider heterostructure scintillator systems, wherein low stopping power nanoplasmonic scintillators are combined with high stopping power materials to combine a higher overall stopping power with nanoplasmonic enhancement in the decay rate.\cite{pagano2022new, turtos2019towards, turtos2019use, lecoq2022metascintillators, pagano2022advances} One could also create multilayers composed of alternating ultrathin scintillator and plasmonic materials to form a scinitllator device of substantial overall thickness, with light extraction from the sides of the structure by arranging to detect waveguided photons or surface plasmon polaritons traveling in the transverse directions. \cite{oulton2008hybrid, liu2005novel} The rich variety of these configurations -- together with the possibility of combining the nanoplasmonic Purcell effect with other advances in scintillator research, such as materials development, dopant engineering and circuit design -- make nanoplasmonic scintillators a fertile and promising area of research towards future heights in ultrafast diagnostics.

In conclusion, we theoretically predict and experimentally demonstrate that plasmonic materials and nanoplasmonic modes can be leveraged to shorten the time resolution and decay time of scintillators through the Purcell effect. Our experiments constitute a key achievement in two respects: Firstly, by demonstrating the Purcell enhancement of scintillators using nanoplasmonics; Secondly, to show that the Purcell effect can be used to further enhance the already ultrafast decay rates (corresponding to $<$2 ns decay times) of scintillator materials like perovskites. To accurate simulate our nanophotonic scintillator system, we develop a theoretical framework that takes the existence of surface plasmon polariton modes fully into consideration. We show that even using a relatively simple, planar design, a nanoplasmonic scintillator can achieve more than 10 times enhancement in decay rate and over 38\% of reduction in detector time resolution can be achieved. In our proof-of-concept experiments for nanoplasmonic enhancement of ultrafast scinitllators -- using perovskite (BA)2PbBr4 as the scintillator and Au as the plasmonic material -- we demonstrate the enhancement of light yield and decay rate of the scintillator by 120\% and 60\%, respectively, in good agreement with the predictions of our theoretical framework. The sub-nanosecond decay time component of 0.7 ns in nanoplasmonic scintillator system shows the possibility to use the scintillator in GHz photon flux pixelated detector. To illustrate the practical utility of the nanoplasmonic scintillator system, we experimentally demonstrate an improved performance. We show that the spatial resolution and contrast preservation of X-ray imaging system can be significantly improved using a nanoplasmonic scintillator device, an advance that is highly promising for various imaging applications such as medical imaging and security inspection. Our findings pave the way to next-generation scintillator devices that combine nanophotonics with other breakthroughs in scintillator development to create highly cost-effective and sensitive detection systems for time-resolved diagnostics in medicine, industry and fundamental research.

\section{\fontsize{12}{20}\selectfont Methods}

\section{\fontsize{12}{20}\selectfont Theoretical Framework for Scintillator process in layered medium with metal interfaces}
In the scintillation process, ionising radiation is converted into numerous electron-hole pairs which will recombine and emit visible photons. These electron-hole pairs can be modelled as decaying dipole emitters with decay rate $\Gamma(\mathbf{p}, \mathbf{r}, \mathbf{k}, \sigma)$, where $\mathbf{p}$ denotes the dipole moment, $\mathbf{r}$ denotes the dipole position, $\mathbf{k}$ denotes the emission wavevector, and $\sigma$ denotes the emission polarisation. 

Under quantum electrodynamics framework, decay rate of dipoles are proportional to the local density of states, i.e., $\Gamma \propto \rho_{\rm p}(\mathbf{p}, \mathbf{r}, \mathbf{k}, \sigma)$. The modification of the decay rate of a dipole emitter is represented by the Purcell factor, i.e.,
\begin{equation}
    \purcell(\theta, z, \lambda, k_{\rho}, \sigma) = \frac{\Gamma(\theta, z, \lambda, k_{\rho}, \sigma)}{\Gamma_0(\lambda)}.
\end{equation}
where $\Gamma_0(\lambda)$ is the total decay rate of a dipole radiating at wavelength $\lambda$ in the free space. The simplest way to modify the local density of states, i.e., to modify decay rate of a dipole emitter is to place the dipole close to a metal interface. \cite{Drx1970,CPR78,Ford1984,Barnes2000} In a one-dimensional layered system (planar interfaces), dipole position $\mathbf{r}$ is denoted by $z$, and emission wavevector $\mathbf{k}$ is denoted by $k_\rho$, with $\hat{\rho}$ and $\hat{z}$ being the radial vector and normal vector, respectively. Here, the dipole moment is assumed to have the same magnitude $p$ and a distribution of orientation $C(\theta)$ intrinsic to the scintillating material, where $\theta$ is the angle between the normal direction and the dipole orientation.

The decay rate $\Gamma(z)$ and emission rate $\Gamma^{\rm T}(z)$ of a dipole oriented perpendicular ($\perp$) or parallel ($\parallel$) to the interfaces at position $z$ could be written as
\begin{equation}
    \Gamma_{\perp}(z) =
    \int \dd \lambda Y(\lambda)
    \int \dd k_{\rho} \frac{k_\rho}{k_{mz} k_m}
    \sum_{\sigma}
    \purcell(\theta=0, z, \lambda, k_{\rho}, \sigma) \Gamma_0(\lambda),
\end{equation}
\begin{equation}
    \Gamma_{\parallel}(z) =
    \int \dd \lambda Y(\lambda)
    \int \dd k_{\rho} \frac{k_\rho}{k_{mz} k_m}
    \sum_{\sigma}
    \purcell(\theta=\pi/2, z, \lambda, k_{\rho}, \sigma) \Gamma_0(\lambda),
\end{equation}
\begin{equation}
    \Gamma^{\rm T}_{\perp}(z) =
    \int \dd \lambda Y(\lambda)
    \int \dd k_{\rho} \frac{k_\rho}{k_{mz} k_m}
    \sum_{\sigma}
    \purcell(\theta=0, z, \lambda, k_{\rho}, \sigma) \Gamma_0(\lambda) T(z, \lambda, k_{\rho}, \sigma),
\end{equation}
\begin{equation}
    \Gamma^{\rm T}_{\parallel}(z) =
    \int \dd \lambda Y(\lambda)
    \int \dd k_{\rho} \frac{k_\rho}{k_{mz} k_m}
    \sum_{\sigma}
    \purcell(\theta=\pi/2, z, \lambda, k_{\rho}, \sigma) \Gamma_0(\lambda) T(z, \lambda, k_{\rho}, \sigma),
\end{equation}
where $Y(\lambda)$ is the intrinsic emission spectrum of the scintillator material, which is normalized by $\int \dd \lambda Y(\lambda) = 1$. $T(z, \lambda, k_{\rho}, \sigma)$ denotes the transmission coefficient of the photons produced from position $z$. $k_m$ and $k_{mz}$ denote the wavenumber and wavenumber in z direction in m$^{\rm th}$ layer, respectively. One key difference from previous study is that we here consider the decay of the dipole in all modes, including outgoing radiative modes, waveguided modes and surface plasmon plariton modes, i.e., $k_{\rho}$ from 0 to $\infty$. Since dipoles can be generated with random orientation in the scintillator, the effective total decay rate of $\Gamma_{\mathrm{eff,tot}}(z)$ and effective total emission rate $\Gamma^{\rm T}_{\mathrm{eff,tot}}(z)$ a dipole located at position $z$ can be written as
\begin{equation}
    \Gamma_{\mathrm{eff,tot}}(z) = \chi \Gamma_{\parallel}(z) + \left[ 1-\chi \right] \Gamma_{\perp}(z).
\end{equation}
\begin{equation}
    \Gamma^{\rm T}_{\mathrm{eff,tot}}(z) = \chi \Gamma^{\rm T}_{\parallel}(z) + \left[ 1-\chi \right] \Gamma^{\rm T}_{\perp}(z).
\end{equation}
where $\chi = \int_{0}^{\pi/2} \dd \theta \sin^{3} \theta C(\theta)$ is the parallel dipole fraction parameter. In this work, we assume an isotropic distribution of dipole orientation ($C(\theta) = 1$ and $\chi = 2/3$). Since the decay rate of a dipole is position-dependent, the averaged dipole decay time of a scintillator system could be written as 
\begin{equation}
    \bar{\tau} = \frac{1}{\bar{\Gamma}} =
    \frac{1}{\int dz G(z) \Gamma_{\mathrm{eff,tot}}(z)}.
\end{equation}
where $G(z)$ denotes the spatial distribution of dipole emitters. In this work, we assume a homogeneous dipole distribution inside scintillator material, i.e., $G(z) = 1/L$, where $L$ is the thickness of the scintillator.

By integrating over the emission angles $\phi_\mathrm{min}$ to  $\phi_\mathrm{max}$, the overall detected emission intensity by a dipole located at position $z$ and perpendicular to the surface is given by
\begin{equation}
    I(z) = I_0 \frac
    {
    \int \dd \lambda Y(\lambda)
    \int_{a}^{b} \dd k_{\rho} \frac{k_\rho}{k_{mz} k_m}
    \sum_{\sigma}
    \Gamma(\theta=0, z, \lambda, k_{\rho}, \sigma)
    T(z, \lambda, k_{\rho}, \sigma)
    }
    {\int \dd \lambda Y(\lambda)
    \int \dd k_{\rho} \frac{k_\rho}{k_{mz} k_m}
    \sum_{\sigma}
    \Gamma(\theta=0, z, \lambda, k_{\rho}, \sigma)},
\end{equation}
where $I_0$ is the total power radiated by a dipole. The limit of integration with respect to $k_\rho$ is $a = k_0 \mathrm{sin} \phi_\mathrm{min}$ and $b = k_0 \mathrm{sin} \phi_\mathrm{max}$, where $k_0 = 2\pi/\lambda$ (we focus on emission intensity in normal direction in this work). The number of photons detected over time combines the influence of emission intensity and the decay rate:

\begin{equation}
    I_{{\rm det}}(t) = N_{\rm d,0} \int dz G(z) I(z) [1-e^{-\Gamma_{\mathrm{eff,tot}}(z) t}],
\end{equation}
where $N_{\rm d,0}$ is the total dipole number in the scintillator. The enhancement of the decay rate, the detected photon number for a scintillator system compared with a bulk scintillator can be written as $I_{{\rm det}}(t=\infty)/I_{{\rm det,Bulk}}(t=\infty)$, and $\bar{\tau}_{{\rm Bulk}}/\bar{\tau}$, respectively. More details can be found in SI Section I.

\section{\fontsize{12}{20}\selectfont Sample Preparation}
We prepared two groups of SiO$_{2}$ substrates for the samples. One group of substrate was used to fabricate perovskite-metal thin film scintillators. The 70 nm thick bottom Au layer was physically deposited on SiO$_{2}$ substrate using electron beam evaporation. To minimize non-radiative losses, a 15-nm thin layer of HfO$_{2}$ was deposited between the Au layer and the perovskite layer. For the perovskite layers, we selected \bapbbr ~and the necessary chemicals provided by Sigma-Aldrich: Dimethyl sulfoxide (DMSO, anhydrous), butylammonium bromide ((BA)Br, 98$\%$), and lead bromide (PbBr$_{2}$, 98$\%$). To obtain a homogeneous solution, the mixture was heated to a temperature exceeding 100 $^{\circ}$C. For the preparation of the film, a precursor solution with varying concentrations ranging from 0.25 to 0.75 M was spin-coated onto the substrates, which were pre-treated with 20 minutes of ultraviolet-ozone exposure (different concentrations resulted in different film thicknesses). The spin coating process involved applying a speed of 8000 rpm for 90 s. After the first 10-s spin coating process, a heat gun was used to blow-dry the film with a flow of hot air for the next 30 s.

\section{\fontsize{12}{20}\selectfont Structure and Composition Characterization}
The structures and the morphologies of perovskite-metal thin film scintillators were characterized using a commercial Helios 660 FEI scanning electron microscope (SEM) (Thermo Fisher, United States) and a Dimension FastScan atomic force microscope (Bruker, United States). The measurements of atomic force microscope were performed using a standard cantilever with a spring constant of 40 N/m and a tip curvature of less than 10 nm. The thicknesses of all perovskite thin films were also checked by Dektak surface profiler (Veeco Instruments, United States). To measure element composition, X-ray photoemission spectroscopy measurements were carried out using an integrated VG ESCA Lab system with a Magnesium K$\alpha$ X-ray source (E$_{\rm ex}$ = 1254 eV) in an ultrahigh vacuum system ($\rho \sim$10$^{-10}$mbar) at room temperature. The size diameter of the impinged spot on the sample was approximately 1 mm.

\section{\fontsize{12}{20}\selectfont PL and TRPL Measurements}
The measurements were conducted in an ambient environment at room temperature using a custom-made micro-photoluminescence setup. A pulsed laser with a wavelength of 355 nm (VisUV, PicoQuant) was employed, which ensures a time resolution of 15 ps and a repetition rate of 10 MHz. The laser beam was focused on the samples using a microscope objective (Olympus, 40× objective, NA = 0.65), resulting in a laser beam diameter of approximately 1 $\upmu {\rm m}$. The filtered PL signal was captured using a thermoelectric-cooled Avaspec HERO spectrometer (Avantes, The Netherlands). For TRPL measurements, the laser repetition rate was reduced to 10 MHz, and the PL signal within a specific range of 400 $\pm$ 25 nm was selectively isolated using a bandpass filter. The selected PL signal was then directed to a single-photon avalanche photodiode (MicroPhoton Devices, Italy). The timing response was analyzed using time-correlated single-photon counting electronics (Becker and Hickl, Germany).

\section{\fontsize{12}{20}\selectfont XL and TRXL Measurements}
For XL measurements, we employed a typical set-up consisting of a home-made X-ray generator (W-anode tube, output energy $\sim$20 keV at operating voltage 33 kV and 2 mA). An Acton SpectraPro-500i monochromator equipped with a 500 nm blazed grating (Princeton Instruments, United States) and a photomultiplier (Hamamatsu, Japan) were also utilized. The XL spectra were collected at an angle of approximately 10$^{\circ}$ from the normal angle and subsequently corrected based on the composition and the thickness of the samples. TRXL measurements were conducted using a time-correlated single photon counting system consisting of a fast photomultiplier tube (SMA650) and PicoHarp 300 electronics (Picoquant, Germany). The excitation source was a N5084 light-excited X-ray tube (Hamamatsu, Japan) operating at 30 kV. Optical excitation of the tube was achieved using a 500kHz Horiba-Delta diode that emits light at 405 nm (DD-405-L). To select the desired emission wavelengths, we employed a Thorlabs FEL450 long-pass interferential filter with a cut-off wavelength greater than 400 nm. The emission was collected at an angle of 45$^{\circ}$ from the normal angle.

\section{\fontsize{12}{20}\selectfont X-ray Imaging Setup}
The X-ray source used in the imaging was the PHYWE XR 4.0 expert unit, equipped with a Cu anode, and operated at 35 kV with a current of 1 mA. For imaging purposes, the camera employed was Alvium with a pixel size of 6.9 $\mu$m (Allied Vision Technologies, Germany). The exposure time for each image was set to be 1 second. A card with a 2.4 mm square hole was positioned at the X-ray tube aperture to allow the uncollimated X-rays to pass through. To minimize additional loss caused by light scattering, we position the card, scintillator, focusing lens, and camera as close as possible to the focal point. To analyze the acquired images, we calculated the edge spread function and its first derivative, commonly referred to as the line spread function. Subsequently, we obtained the modulation transfer function by taking the modulus of the Fourier transform of the line spread function.

\section{\fontsize{12}{20}\selectfont Acknowledgements}
The authors are grateful for the support of Wesley Lee Wei Wong in graphics preparation. L.J.W. acknowledges the Nanyang Assistant Professorship Start-up Grant. DK and M.D.B. acknowledge funding from National Science Center, Poland under grants MINIATURA no. 2022/06/X/ST5/00369 and OPUS-24 no. 2022/47/B/ST5/01966.

\section{\fontsize{12}{20}\selectfont Supporting Information}

Supporting Information is available from the author.

\bibliography{achemso-demo}
\end{document}